\documentclass[10pt,aps,prl,twocolumn,nopacs,final,letterpaper,superscriptaddress,longbibliography]{revtex4-1}

\usepackage[utf8]{inputenc}
\usepackage{calc}
\usepackage{graphicx}
\usepackage{amsmath,amssymb,amsthm}
\usepackage{mathtools}
\usepackage{txfonts}
\usepackage{bm}
\usepackage{color}
\usepackage{hyperref}
\usepackage{multirow}
\graphicspath{{./figs/}}

\begin{document}
\author{Guillaume St-Onge}
\affiliation{D\'epartement de physique, de g\'enie physique et d'optique,
Universit\'e Laval, Qu\'ebec (Qu\'ebec), Canada G1V 0A6}
\affiliation{Centre interdisciplinaire en mod\'elisation math\'ematique, Universit\'e Laval, Qu\'ebec (Qu\'ebec), Canada G1V 0A6}
\author{Vincent Thibeault}
\affiliation{D\'epartement de physique, de g\'enie physique et d'optique,
Universit\'e Laval, Qu\'ebec (Qu\'ebec), Canada G1V 0A6}
\affiliation{Centre interdisciplinaire en mod\'elisation math\'ematique, Universit\'e Laval, Qu\'ebec (Qu\'ebec), Canada G1V 0A6}
\author{Antoine Allard}
\affiliation{D\'epartement de physique, de g\'enie physique et d'optique,
Universit\'e Laval, Qu\'ebec (Qu\'ebec), Canada G1V 0A6}
\affiliation{Centre interdisciplinaire en mod\'elisation math\'ematique, Universit\'e Laval, Qu\'ebec (Qu\'ebec), Canada G1V 0A6}
\author{Louis J. Dub\'{e}}
\affiliation{D\'epartement de physique, de g\'enie physique et d'optique,
Universit\'e Laval, Qu\'ebec (Qu\'ebec), Canada G1V 0A6}
\affiliation{Centre interdisciplinaire en mod\'elisation math\'ematique, Universit\'e Laval, Qu\'ebec (Qu\'ebec), Canada G1V 0A6}
\author{Laurent H\'{e}bert-Dufresne}
\affiliation{D\'epartement de physique, de g\'enie physique et d'optique,
Universit\'e Laval, Qu\'ebec (Qu\'ebec), Canada G1V 0A6}
\affiliation{Department of Computer Science \& Vermont Complex Systems Center, University of Vermont, Burlington, VT 05405 }

\title{Social confinement and mesoscopic localization of epidemics on networks}

\begin{abstract} 
Recommendations around epidemics tend to focus on individual behaviors, with much less efforts attempting to guide event cancellations and other collective behaviors since most models lack the higher-order structure necessary to describe large gatherings. Through a higher-order description of contagions on networks, we model the impact of a blanket cancellation of events larger than a critical size and find that epidemics can suddenly collapse when interventions operate over groups of individuals rather than at the level of individuals. We relate this phenomenon to the onset of mesoscopic localization, where contagions concentrate around dominant groups.
\end{abstract}

\maketitle

Standard disease models reduce the complexity of epidemics to simple processes that provide useful insights. In fact, many of the key results of these models provide the foundation for our current understanding and forecasting of novel emerging epidemics \cite{KMK1, KMK2, KMK3}. The reduction of the complex to the simple is perhaps best embodied by the \textit{mass-action approximation} \cite{Diekmann1995}. This assumption essentially means that we are considering a randomly mixed population, ignoring household structures, social gatherings, and the different behaviors of different individuals.
Mass-action models are thus seriously limited, since they focus on the average number of infections caused by each case, the basic reproduction number $R_0$ \cite{anderson1992infectious}, and ignore the underlying heterogeneity \cite{Hebert-Dufresne2020a}.
There are also conceptual issues with the design of targeted interventions when relying on the mass-action assumption. Where should we target our interventions, and what should be their impact? In this letter, we address these questions through higher-order contact patterns.

Network science provides a natural framework to go beyond the mass-action approximation by considering key features of the structure of contacts among individuals.
The simplest generalization is perhaps the \textit{heterogeneous pair approximation}---individuals are nodes categorized by their number of contacts and their state, and the contacts are distinguished by the states of the nodes involved \cite{Pastor-Satorras2015}.
At this level of sophistication, all pairwise contacts of a given state are still, a priori, equivalent.

\begin{figure}[tb]
\centering
\includegraphics[width=0.98\linewidth]{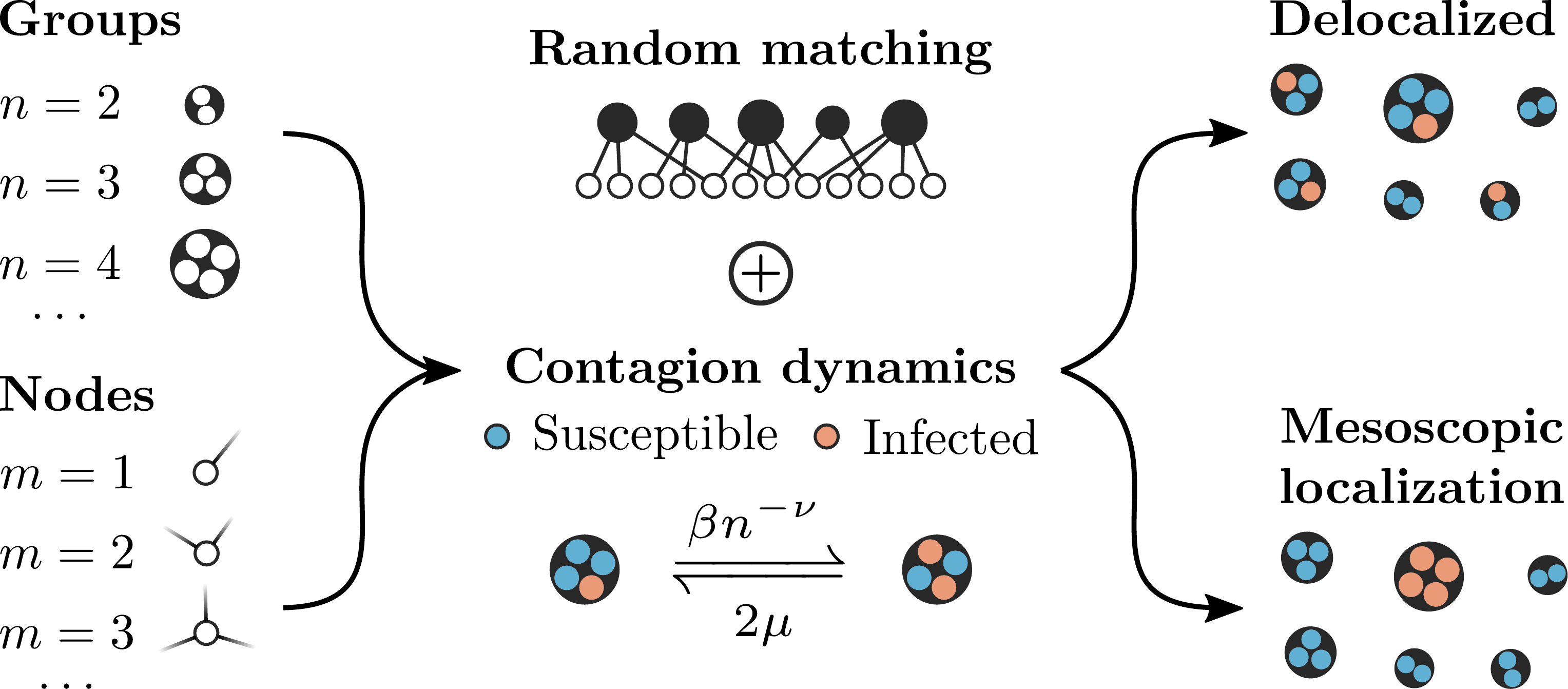}
\caption{\textbf{Framework for contagions on higher-order networks.}
Nodes are assigned to $m$ groups and groups are of various sizes $n$, distributed according to $g_m$ and $p_n$. We consider a SIS dynamics where infected nodes transmit the disease in a group of size $n$ at rate $\beta n^{-\nu}$ with $\nu \in [0,1]$, and recover at rate $\mu$. We characterize the phenomenon of mesoscopic localization, namely the concentration of infected nodes in large groups.}
\label{fig:framework}
\end{figure}

\begin{figure*}[tb]
  \centering
  \includegraphics[width=0.98\linewidth]{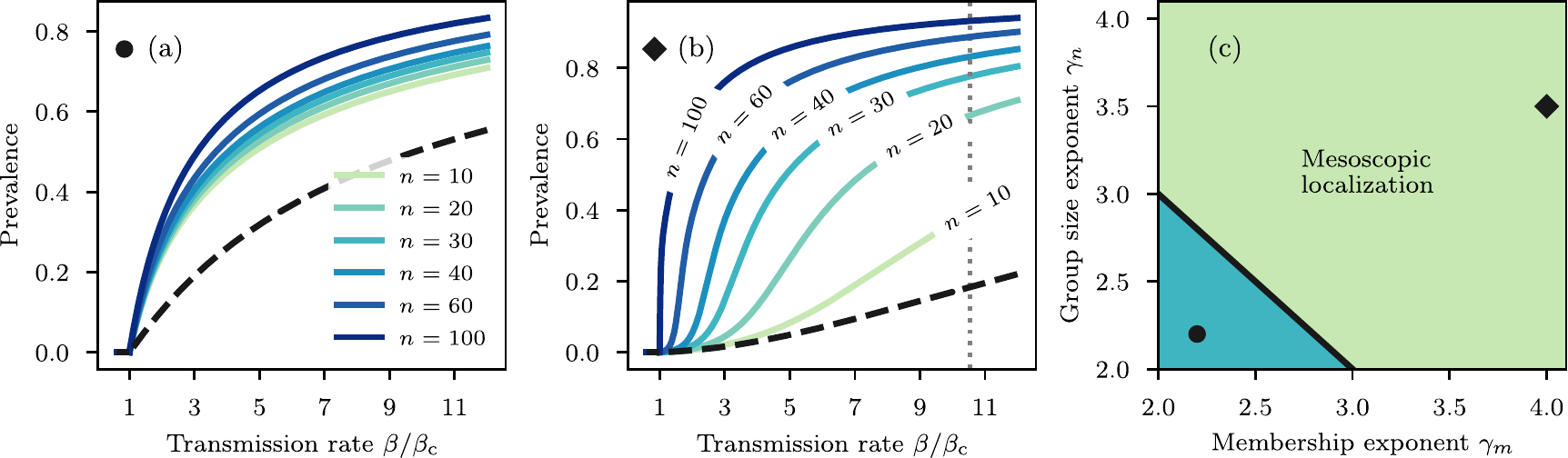}
  \caption{\textbf{Epidemic localization in networks with heterogeneous higher-order structures.} We use $\nu = 0$ for all panels and power-law distributions for the memberships $g_m \propto m^{-\gamma_m}$ and group sizes $p_n \propto n^{-\gamma_n}$. (a)-(b) Solid lines represent the stationary group prevalence and dashed lines represent the stationary global prevalence.
  (a) For strongly coupled groups $(\gamma_m = \gamma_n = 2.2)$, we find a collective phase transition.
  (b) For weakly coupled groups $(\gamma_m = 4, \gamma_n = 3.5)$, we find a phenomenon of mesoscopic localization. While the global prevalence in the population can remain extremely low, larger groups can self-sustain the epidemic. The vertical dotted line is an estimation of the \textit{delocalization} threshold, where the epidemic invades the whole network (see Ref.~\cite{PRE}).
  (c) Mesoscopic localization is the rule rather than the exception.  The solid line, $\gamma_m + \gamma_n =5$, separates the \textit{delocalized regime} (blue area of strong group coupling) from the \textit{mesoscopic localization regime} (green area of weak group coupling), obtained from Eqs.~(\ref{eq:localization_condition}a-b). The circle and diamond markers correspond to the networks in panels (a)-(b).}
  \label{fig:local}
\end{figure*}

One developing area in network science concerns dynamical processes on higher-order representation of networks, i.e., where the network is not simply a conglomerate of pairwise interactions but where interactions occur in a coordinated manner because of a higher-level organization (schools, households, events, etc.) \cite{Battiston2020}.
For dynamics on higher-order networks, one straightforward generalization of the framework just described is the \textit{heterogeneous clique approximation} \cite{hebert2010propagation}. Nodes are categorized by their state and \textit{membership}, i.e., the number $m$ of groups to which they belong. The groups are characterized by their size $n$, and the states of the nodes involved.

Let us consider contagion processes on a simple version of higher-order networks, see Fig.~\ref{fig:framework}.
The network is characterized by $g_m$, the distribution for the memberships $m$ of nodes and $p_n$, the distribution for the sizes $n$ of groups. We use different heterogeneous distributions for both of them, $g_m \propto m^{-\gamma_m}$ and $p_n \propto n^{-\gamma_n}$, with finite cut-offs $m_\mathrm{max}$ and $n_\mathrm{max}$ respectively.

For mathematical convenience, we use a Susceptible-Infected-Susceptible (SIS) dynamics. However, our results have repercussions for a much broader class of dynamical processes, including Susceptible-Infected-Recovered (SIR) dynamics where individuals develop immunity.
Infected nodes transmit the disease to susceptible nodes belonging to a same group of size $n$ at rate $\beta n^{-\nu}$ with $\nu \in [0,1]$, and recover at rate $\mu$.
It is equivalent to a standard SIS model on networks formed of cliques \cite{Newman2003,hebert2010propagation}, but the edges have weights $n^{-\nu}$. We recover an unweighted network for $\nu = 0$.
The parameter $\nu$ tunes the strength of interactions within groups, which we assume to be decreasing with size. For instance, an individual in a workplace typically interacts with more people than at home, but interactions in a household are stronger.

We track $s_m(t)$, the probability for a node of membership $m$ to be susceptible, and $c_{n,i}(t)$, the probability to observe $i$ infected nodes within a group of size $n$. Their dynamics are described by the following coupled ordinary differential equations \cite{hebert2010propagation},
\begin{subequations}
\label{eq:ode}
\begin{align}
    \frac{\mathrm{d}s_m}{\mathrm{d}t} =& \;\mu (1 - s_m) - m r s_m  \;, \\
    \frac{\mathrm{d}c_{n,i}}{\mathrm{d}t} =& \;\mu (i+1) c_{n,i+1} - \mu i c_{n,i} +  (n-i+1) \lbrace \beta n^{-\nu}(i-1) + \rho\rbrace c_{n,i-1} \notag  \\
    &-  (n-i) \lbrace \beta n^{-\nu} i + \rho\rbrace c_{n,i} \;, \label{eq:ame}
\end{align}
\end{subequations}
with the mean fields $r(t)$ and $\rho(t)$ defined as
\begin{subequations}
\begin{align}
    r(t) &= \frac{\sum_{n,i}\beta n^{-\nu} i \;(n-i)c_{n,i}(t)p_n}{\sum_{n,i} (n-i) c_{n,i}(t)p_n} \;, \\
    \rho(t) &= r(t)\frac{\sum_{m} (m-1) \; m s_m(t) g_m }{\sum_{m} m s_m(t) g_m} \;.
\end{align}
\end{subequations}
If we take a susceptible node and select a random group to which it belongs, $r(t)$ is the mean infection rate associated to that group.
Now if we pick a susceptible node in a group, $\rho(t)$ is the mean infection rate received from all external groups (i.e., excluding the one we picked the node from).
Without loss of generality, we set $\mu = 1$ hereafter.

An important feature of this framework is that Eq.~(\ref{eq:ame}) is an \textit{approximate master equation} : it describes the full range of possible states for groups of size $n$, while assuming a mean-field coupling between them. As we show in Ref.~\cite{PRE}, the agreement with Monte-Carlo simulations is excellent.
The global prevalence in the network---the average fraction of infected nodes---is then
\begin{align}\label{eq:global_prevalence}
    I(t) = \sum_m [1 - s_m(t)]\; g_m \;,
\end{align}
and the group prevalence is
\begin{align}\label{eq:group_prevalence}
    I_n(t) = \sum_i \frac{i}{n} \; c_{n,i}(t)\;.
\end{align}

In Fig.~\ref{fig:local}(a) and Fig.~\ref{fig:local}(b), we show the stationary prevalence (global and within groups) for two different networks, obtained using Eq.~\eqref{eq:ode}.
As expected from standard models, there exists an epidemic threshold $\beta_\mathrm{c}$ for the transmission rate below which epidemics cannot be sustained (see Ref.~\cite{PRE} for an analytical expression of $\beta_\mathrm{c}$).
Above $\beta_\mathrm{c}$, the disease-free equilibrium of the dynamics becomes unstable, driving the epidemic to invade the network.

What is less expected are the sequential local transitions observed in the second panel [Fig.~\ref{fig:local}(b)]. For any value of the transmission rate, the outbreak thrives only in groups above a certain size. The epidemic is \textit{self-sustained} locally, and the global prevalence reaches its highest growth rate with $\beta$ well above the epidemic threshold, a defining feature of \textit{smeared} phase transitions \cite{Vojta2006,hebert2019smeared}.
This is reminiscent of certain infections, such as the bacteria  \textit{C. difficile}, mainly found in hospitals with large susceptible populations in close contact \cite{mcfarland1986review}.

To get some insights on the emergence of this localization phenomenon, we examine the stationary group prevalence, $I_n^*$,  near the absorbing-state. Using a saddle-point approximation valid for large $n$, we obtain \cite{PRE}
\begin{align}
    \label{eq:asymptotic_In}
    I_n^* \sim \begin{dcases}
        \frac{1}{1 - \beta n^{1-\nu}} & \text{if } \beta < n^{\nu-1} \\
        n^{1/2} \;(\beta  n^{1-\nu})^{n} \;e^{-n+n^\nu/\beta } & \text{if } \beta \geq n^{\nu-1} \;.
    \end{dcases}
\end{align}
For $\beta > n^{\nu-1}$, this implies $I_n^* = \mathcal{O}\left (n^{1/2}e^{bn} \right)$ with \mbox{$b > 0$}.
Therefore, if $\beta_\mathrm{c} \to n_\mathrm{max}^{\nu-1}$, the group prevalence increases exponentially with $n$ above the epidemic threshold, and the outbreak is \textit{localized} in large groups, as observed in Fig.~\ref{fig:local}(b).
In other words, the behavior of the epidemic threshold dictates whether or not localization is possible for a given network organization.

In Ref.~\cite{PRE}, we show that for power-law distributions of membership and group size, we have the following behavior
\begin{align}\label{eq:asymptotic_threshold}
    \beta_\mathrm{c}^{-1} \sim \Omega(g_m,p_n ; \nu) + n_\mathrm{max}^{1-\nu} \;,
\end{align}
where we define the \textit{coupling} between groups as
\begin{align}
    \label{eq:coupling}
    \Omega(g_m,p_n ; \nu) = \left (\frac{\langle m(m-1) \rangle}{\langle m \rangle} \right) \left ( \frac{ \left \langle n^{1-\nu} (n-1) \right \rangle}{ \langle n \rangle} \right) \;.
\end{align}
Asymptotic analysis of Eqs.~\eqref{eq:asymptotic_threshold} and \eqref{eq:coupling} in the limit $n_\mathrm{max} \to \infty$ reveals the conditions for which $\beta_\mathrm{c} \to n_\mathrm{max}^{\nu-1}$, i.e., the conditions necessary for a localized epidemic \cite{PRE}.
They require $\nu < 1$, and are always met whenever $\gamma_m \geq 3$. If $2 < \gamma_m < 3$, the conditions are then satisfied only if
\begin{subequations}
\label{eq:localization_condition}
\begin{align}
    &2 < \gamma_n + \nu < 3 \; \text{ and } \; 3-\gamma_n + \alpha(3-\gamma_m) < 1 \;, \\
    \text{or} \quad \quad & \notag \\
    & \gamma_n + \nu \geq 3 \; \text{ and } \; \alpha(3-\gamma_m) + \nu < 1 \;,
\end{align}
\end{subequations}
where $\alpha \geq 0$ relates the two cut-offs $m_\mathrm{max} \sim n_\mathrm{max}^\alpha$.
These conditions define \textit{mesoscopic localization} and distinguish the \textit{localized} regime from the \textit{delocalized} regime.
We give some examples with $\nu = 0$ and $\alpha = 1$ in Fig.~\ref{fig:local}(c).

More intuitively, the conditions (\ref{eq:localization_condition}a-b) can be interpreted as the result of a competition between a collective and a local activation \cite{Ferreira2016}.
All pairs of $(\gamma_m, \gamma_n)$ below the dark dividing line in Fig.~\ref{fig:local}(c) are associated with a strong group coupling ($\Omega \gg n_\mathrm{max}^{1-\nu}$), whereas pairs above the line correspond to a weak group coupling ($\Omega \ll n_\mathrm{max}^{1-\nu}$).

One important observation is that a large fraction of the structural parameter space $(\gamma_m, \gamma_n)$ corresponds to the mesocopic localization regime [green region in Fig.~\ref{fig:local}(c)], making it the rule rather than the exception.
Moreover, the delocalized regime [blue region in Fig.~\ref{fig:local}(c)] is the parameter sub-space where the underlying networks are dense, i.e, where the average number of contacts of a node, proportional to $\left \langle  n(n-1) \right \rangle$, diverges in the asymptotic limit.
Since real-world networks are generally sparse \cite{DelGenio2011,newman2018networks}, it is reasonable to assume mesoscopic localization may occur in many real-world networks with a higher-level organization.
The results of Fig.~\ref{fig:local} extend nicely to cases with $\nu > 0$ (see Ref.~\cite{PRE}, Appendix E) where the localized regime still dominates in a large portion the structural parameter space.

It is worth noting that we only observe localization for a certain portion $\beta \in [\beta_\mathrm{c}, \beta^*]$ of the bifurcation diagram, where $\beta^*$ is the \textit{delocalization threshold} [dotted line in Fig.~\ref{fig:local}(b)].
In Ref.~\cite{PRE}, we show how to estimate $\beta^*$ and discuss other details of the localization regimes, notably the effects of finite cut-offs.

Perhaps most surprising about mesoscopic localization is how strong the effect can be. Even at low overall prevalence, we observe intense but local outbreaks in large groups.
This simple observation justifies targeted interventions on these groups, analogous to school closures and the cancellation of large social or professional events.
While such closures may seem excessive given the low prevalence found in the general population, these compact organized groups are where most infections will occur.

We now focus on targeted interventions on large groups, modeled by simply forcing a hard cut-off $n_\mathrm{max}$ on the distribution $p_n$.
By using a cut-off instead of immunizing the groups, we preserve the membership $m$ of nodes; they will simply belong to groups of smaller sizes.

\begin{figure}[tb]
\centering
\includegraphics[width=\linewidth]{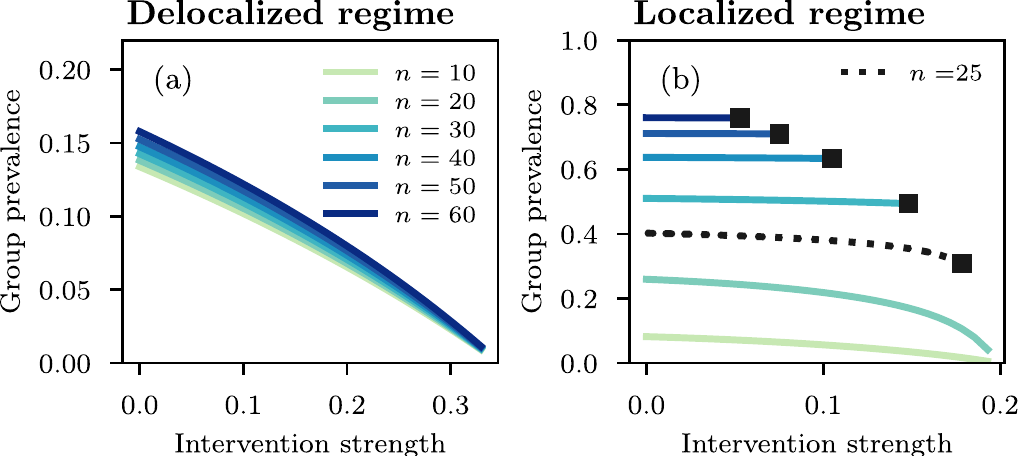}
\caption{\textbf{Local impact of structural interventions in delocalized and localized epidemics.} (a)-(b) Prevalence within groups of size $n$ against the intervention strength [Eq.~\eqref{eq:intervention_strength}]. We use the networks from Fig~\ref{fig:local}(a-b) with transmission rates adjusted to have similar global prevalence for both regimes without intervention. (a) In the delocalized regime, using $\beta \approx 0.0041$, we find a similar benefit of the intervention among all groups. (b) In the mesoscopic localization regime, using $\beta = 0.07$, we find a different story. Large groups that have not been removed by the intervention are barely affected, until the intervention is strong enough to cause a global collapse of the epidemic. Square markers indicate when groups of a particular size are removed.}
\label{fig:inter_local}
\end{figure}

To compare the effectiveness of interventions across different networks and parametrizations, we define an \textit{intervention strength} (IS) as the fraction of the total edge weights that have been removed.
If $p_n$ and $\tilde{p}_n$ are respectively the group size distribution before and after the intervention, then
\begin{align}
    \label{eq:intervention_strength}
    \text{IS} = 1 - \frac{\sum_{n} n^{1-\nu}(n-1)\tilde{p}_n}{\sum_{n} n^{1-\nu}(n-1) p_n} \;.
\end{align}

In Fig.~\ref{fig:inter_local}, we show the local impact on the group prevalence for such structural interventions.
For networks in the delocalized regime [Fig.~\ref{fig:inter_local}(a)], the intervention appears to reduce the risk of infections at the individual (node) level. As we decrease $n_{\textrm{max}}$ (and therefore increase the intervention strength), the local prevalence within all groups decreases gradually and homogeneously until an epidemic threshold is reached.
This is similar to traditional models where interventions reduce $R_0$ in a distributed, mass-action way.

In the localized regime [Fig.~\ref{fig:inter_local}(b)], the intervention has a very different impact. Individuals that would have interacted in groups of size greater than $n_{\textrm{max}}$ are spared by the intervention, but the large groups of sizes below $n_{\textrm{max}}$ appear unaffected by the intervention.
The main point here is that a local outbreak in certain organized gatherings (e.g. mass transit in urban centers, cruise ships) can persist despite interventions elsewhere.
However, once the intervention is strong enough [roughly when $n_\mathrm{max}$ is below $25$ in Fig.~\ref{fig:inter_local}(b)], further interventions cause a rapid collapse of the epidemic.

\begin{figure}[tb]
\centering
\includegraphics[width=\linewidth]{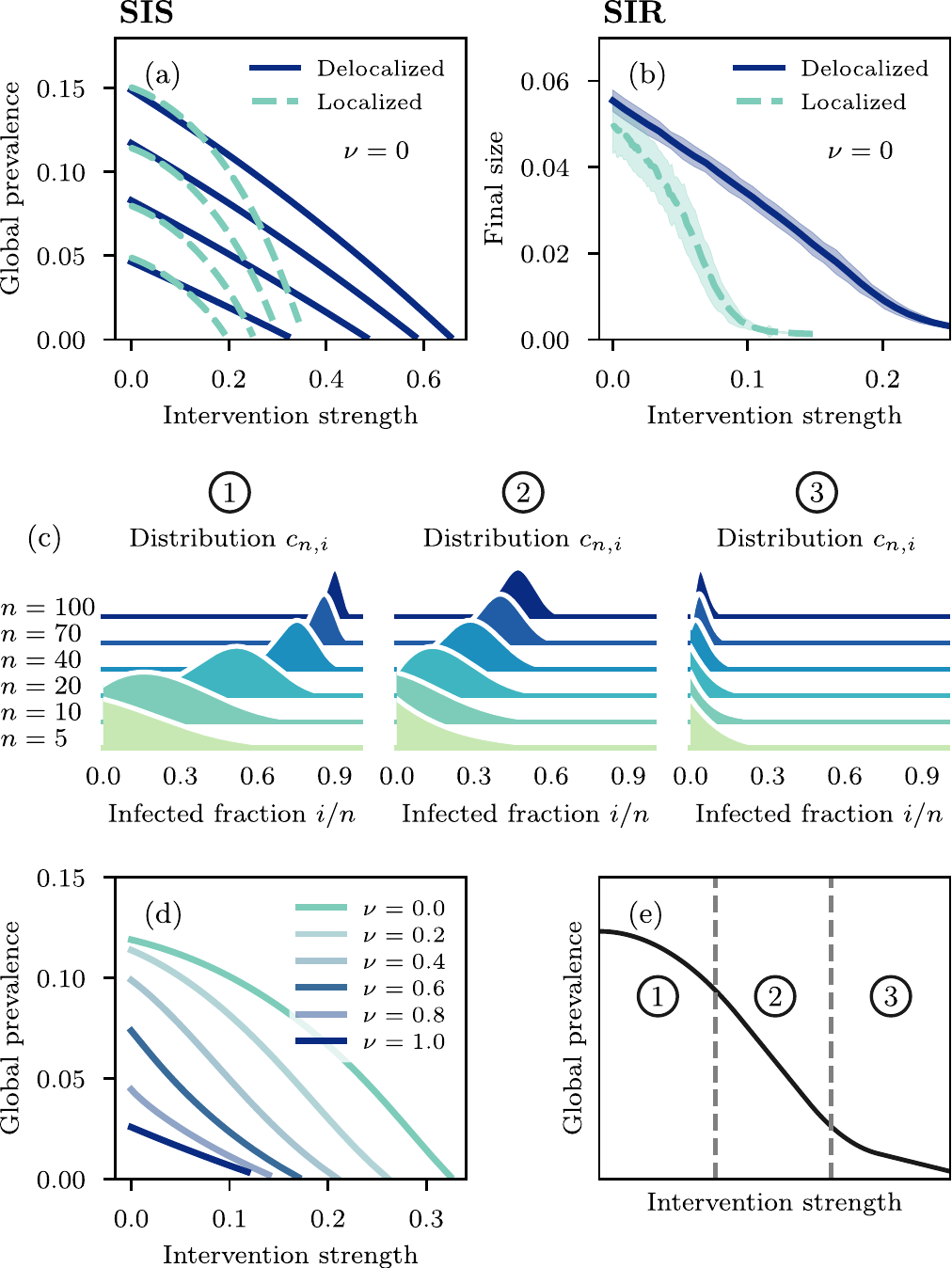}
\caption{\textbf{Global impact of structural interventions in different localization regimes of SIS and SIR dynamics.} (a)~We compare the global prevalence of epidemics against the intervention strength [Eq.~\eqref{eq:intervention_strength}] using the same networks as in Fig.~\ref{fig:inter_local}. We use $\beta \in \lbrace 0.004, 0.005, 0.006, 0.007 \rbrace$ and $\beta \in \lbrace 0.07,0.09,0.11,0.13\rbrace$ for the delocalized and localized regimes. (b)~A similar behavior is observed for the final size (fraction of recovered nodes at $t \to \infty$) in simulations of the SIR dynamics \cite{St-Onge2019}. We generate 1000 networks of size $N = 10^6$ and run $10^4$ SIR simulations on each network, starting with a single random infected node, with $\beta = 0.0036$ in the delocalized regime and $\beta = 0.083$ in the localized regime. We only kept macroscopic outbreaks, with a final size over $10^{-3}$. Solid and dashed lines are sample means and the shaded regions represent twice the standard deviation. (c)~Three examples of $c_{n,i}$ distributions for (1) strongly localized, (2) weakly localized, and (3) delocalized epidemics, using $\nu$ equal 0, 0.6 and 1 respectively. We used networks with $\gamma_m = \gamma_n = 3.5$ and $\beta = 0.3/\langle k \rangle$, where $\langle k \rangle$ is the average weighted degree before interventions. (d)~We look at the impact of interventions for intermediate localization regimes by varying $\nu$ with the same networks as in (c). (e)~Schematic representation of how the impact of interventions varies as it passes through different localization regimes.}
\label{fig:inter_global}
\end{figure}

Let us now look at the global impact of these interventions in Fig.~\ref{fig:inter_global}.
It is important to note that while we have assumed a SIS dynamics, our results hold for a SIR dynamics as well [see the similarity of Fig.~\ref{fig:inter_global}(a-b)], which is generally a more realistic model for epidemics.
The reason is that smeared phase transitions occur for this type of process as well \cite{hebert2019smeared}.

In Fig.~\ref{fig:inter_global}(a-b), we observe that the global impact of interventions in the delocalized regime is again similar to mass-action models---the prevalence decreases approximately linearly with the intervention strength.
In contrast, in the localized regime, the intervention has a rapid non-linear effect in reducing the global prevalence as its strength is ramped up.

In both Fig.~\ref{fig:inter_local}(b) and Fig.~\ref{fig:inter_global}(a-b), it is surprising that the removal of the largest groups does not produce the largest decrease in prevalence, since these are the ones in which we expect most nodes to be infected.
Even less clear is what drives the sudden collapse and if we should expect this behavior for all localized epidemics.

To clarify the situation, we need to explore the regimes between the delocalized and the \textit{strongly} localized regimes displayed in Fig.~\ref{fig:inter_local}(b) and \ref{fig:inter_global}(a-b).
We break down these regimes in three parts in Fig.~\ref{fig:inter_global}(c) to simplify the discussion: (1) strongly localized, (2) weakly localized, and (3) delocalized.
In the strongly localized regime (1), large groups have a high prevalence, and act as independent entities that are barely affected by interventions elsewhere [as in Fig.~\ref{fig:inter_local}(b)].
In the weakly localized regime (2), the disease still thrives in large groups, but they are not isolated from one another---interventions in one group now affect the others as well. This is the regime where interventions are most effective.
Finally, in the delocalized regime (3), groups act as a whole, but the infection does not thrive in any of them, leading to a lower effectiveness of targeted interventions.

The easiest way to interpolate between the two extremes is to tune the group interaction strength through $\nu$.
As shown in Fig.~\ref{fig:inter_global}(d), the initial impact of interventions changes as $\nu$ is increased.
With $\nu = 0.6$ for instance, the epidemic is weakly localized, and now the removal of the largest groups produce the largest decrease in prevalence.
Further interventions become eventually less effective, as the intervention itself causes the epidemic to shift to a delocalized regime.

More generally, the non-linear decrease of the global prevalence as a function of the intervention strength is explained by transitions between different localization regimes, as illustrated in Fig.~\ref{fig:inter_global}(e).
The sudden collapse for interventions on strongly localized epidemics is thus the result of a shift to a weakly localized regime, where interventions become much more effective.

Altogether, the lesson from Fig.~\ref{fig:inter_local} and \ref{fig:inter_global} is that, just as we take heterogeneity of individual risks into account when preferentially vaccinating individuals, we should take heterogeneity of group risks into account when designing interventions.
While pathogens operate at the scale of individuals, epidemics themselves interact with our entire social network, which has a modular, hierarchical, higher-order structure.
Since we expect real epidemics to experience localization effects, we should aim to leverage their sudden collapse when designing structural interventions, and account for this emergent feature in our models.

Over the last few years, dynamics on higher-order representation of networks have shown time and time again that intuition built from simpler models does not always hold in more complex scenarios, with examples ranging from competitive dynamics \cite{grilli2017higher} to social contagion \cite{iacopini2019simplicial}. Localization of dynamics over higher-order structures is yet another addition to this list. Our work thus paves the way for a more comprehensive analysis of structural interventions on networks.

\section*{Acknowledgments}
The authors acknowledge Calcul Qu\'{e}bec for computing facilities.
L.H.-D. was supported by the National Institutes of Health 1P20 GM125498-01 Centers of Biomedical Research Excellence Award.
This work was also supported by the Fonds de recherche du Qu\'{e}bec – Nature et technologies (V.T., G.S.), the Natural Sciences and Engineering Research Council of Canada (G.S., V.T., A.A., L.J.D.), and the Sentinelle Nord program of Universit\'{e} Laval, funded by the Canada First Research Excellence Fund (G.S., V.T., A.A., L.J.D.).
L.H.-D. also acknowledges the dedication of his ALife 2020 co-organizers, Juniper Lovato and Josh Bongard, for moving the conference online to reduce the spread of COVID-19.
The authors thank Simon DeDeo who inspired them to write this letter.


%

\end{document}